\documentclass[12pt]{article}
\usepackage{epsfig,latexsym,amssymb,amsmath}
\usepackage{color}
\usepackage{graphicx}
\usepackage[linktocpage,pagebackref]{hyperref}

\begin{document}

\begin{center}
{\large \bf Thermodynamic Product Formula for Taub-NUT  Black Hole}
\end{center}

\vskip 5mm

\begin{center}
{\Large{Parthapratim Pradhan\footnote{E-mail: pppradhan77@gmail.com}}}
\end{center}

\vskip  0.5 cm

{\centerline{\it Department of Physics}}
{\centerline{\it Vivekananda Satavarshiki Mahavidyalaya}}
{\centerline{\it (Affiliated to Vidyasagar University)}}
{\centerline{\it Manikpara, West Midnapur}}
{\centerline{\it West Bengal~721513, India}}

\vskip 1cm

\begin{abstract}
We derive various important thermodynamic relations of the inner and outer horizon in the 
background of Taub-NUT(Newman-Unti-Tamburino) black hole in four dimensional 
\emph{Lorentzian geometry}. We compare these properties with the properties of Reissner 
Nordstr{\o}m black hole. We compute \emph{area product, area sum, area minus and area division} 
of black hole horizons. We show that they all are not universal quantities. Based on these relations, we 
compute the area bound  of all horizons. From area bound, we derive entropy bound and irreducible 
mass bound for both the horizons. We further study the stability of such black hole by computing 
the specific heat for both the horizons. It is shown that due to negative specific heat the 
black hole is thermodynamically unstable. All these calculations might be helpful to understanding 
the nature of black hole entropy both \emph{interior} and exterior at the microscopic level. 
\end{abstract}

%\begin{keyword}
%Area product, Entropy product, Entropy bound, Area bound, Taub-NUT Black Hole.
%\end{keyword}

\section{Introduction}
There has recently been intense interest in the thermodynamic product formulae in both 
general relativity community\cite{ah09} and string/M-theory community\cite{cgp11} to understand 
the black hole(BH) entropy\cite{bk72} of multi-horizons at the microscopic level.  Significant 
achievements have been made in case of asymptotically flat super-symmetric BHs in four and five 
dimension, where the microscopic degrees of freedom could be explained in terms of a 
two-dimensional(2D) conformal field theory(CFT)\cite{vapa96}. For microscopic entropy of 
extreme rotating BH the results could be found in \cite{ghss09}.

In case of a regular axisymmetric and stationary spacetime of Einstein-Maxwell gravity 
with surrounding matter, the area product formula of event horizon(${\cal H}^{+}$)
and Cauchy horizons(${\cal H}^{-}$) is\cite{ah09}
\begin{eqnarray}
\frac{{\cal A}_{+} {\cal A}_{-}}{64\pi^2} &=& J^2+\frac{Q^4}{4} ~.\label{prKN}
\end{eqnarray}
and consequently the entropy product formula of ${\cal H}^\pm$ is
\begin{eqnarray}
\frac{{\cal S}_{+} {\cal S}_{-}}{4\pi^2} &=& J^2+\frac{Q^4}{4} ~.\label{prKN1}
\end{eqnarray}
In the absence of Maxwell gravity, these product formulae reduces to the following 
form\cite{ac79}:
\begin{eqnarray}
\frac{{\cal A}_{+} {\cal A}_{-}}{64\pi^2} &=& J^2 ~.\label{prK}
\end{eqnarray}
and
\begin{eqnarray}
\frac{{\cal S}_{+} {\cal S}_{-}}{4\pi^2} &=& J^2 ~.\label{prK1}
\end{eqnarray}

In the above formulae, the interesting point is that they all are independent of the mass, so-called the
ADM(Arnowitt-Deser-Misner) mass of the back-ground space-time. Thus they all are universal quantities 
in this sense. Now if we incorporate the BPS states, the area product formula should be\cite{cgp11}
\begin{eqnarray}
\frac{{\cal A}_{+} {\cal A}_{-}}{64\pi^2}  &=& \left(\sqrt{N_{1}}\pm\sqrt{N_{2}}\right)
= N , \,\, N\in {\mathbb{N}}, N_{1}\in {\mathbb{N}}, N_{2} \in {\mathbb{N}} ~.\label{ppl}
\end{eqnarray}
where the integers $N_{1}$ and $N_{2}$ may be viewed as the excitation numbers of the left and 
right moving modes of a weakly-coupled two-dimensional CFT. $N_{1}$ and $N_{2}$ are depends 
explicitly on all the BH parameters.

Consequently, the entropy product formula of ${\cal H}^{\pm}$ should be 
\begin{eqnarray}
\frac{{\cal S}_{+} {\cal S}_{-}}{4\pi^2}  &=&  \left(\sqrt{N_{1}}\pm\sqrt{N_{2}}\right)
= N , \,\, N\in {\mathbb{N}}, N_{1}\in {\mathbb{N}}, N_{2} \in {\mathbb{N}} ~.\label{ss}
\end{eqnarray}
It implies that the product of ${\cal H}^{\pm}$ could be expressed in terms of the underlying 
CFT which can be  also interpreted in terms of a level matching condition. That means the 
product of entropy of ${\mathcal H}^{\pm}$ is an integer quantity\cite{fl97}.

It is true that certain BH possesses inner horizon or Cauchy horizon(CH) in addition to the
outer horizon or event horizon. Thus there might be a relevance of inner horizon in BH thermodynamics
to understanding the microscopic nature of inner BH entropy in comparison with the outer BH entropy. It is
also true that CH is a blue-shift region whereas event horizon is a red-shift region by its own nature.
Furthermore the CH is highly unstable due to the exterior perturbation\cite{sc83}. Despite the above
characteristics, the CH horizon is playing a crucial role in BH  thermodynamics.

In this work, we shall focus on thermodynamic properties of both inner horizon and outer horizons 
of Lorentzian Taub-NUT BH. The special property of this BH is it is non-asymptotic type in comparison 
with RN BH, which is asymptotic type. The special features give a motivation to study them.

The plan of the paper is as follows.
In Sec.(\ref{tnt}), we describe various thermodynamic properties for Taub-NUT BH. In this
Sec., there are five sub-section. In first sub-section, we derive the area bound of all horizons for 
Taub-NUT BH. In second sub-section, we derive the entropy bound for Taub-NUT BH. In third subsection, we compute 
the irreducible mass bound of ${\cal H}^{\pm}$. In fourth subsection, we derive temperature bound of all horizons 
and finally in last sub-section, we compute specific heat bound for both the horizons.
Finally, we conclude in Sec.(\ref{dis}).

\section{\label{tnt} Thermodynamic properties of Lorentzian Taub-NUT BH:}
The Lorentzian Taub-NUT(TN) BH is a stationary, spherically-symmetric vacuum solution of Einstein 
equations with the NUT parameter $(n)$. The NUT charge or dual mass has an intrinsic feature 
in Einstein's  general relativity and which is a gravitational analogue of a magnetic monopole 
in Maxwell's electrodynamics \cite{nouri98}. The presence of the NUT parameter in the space-time
destroy its asymptotic structure making it, in contrast to the RN space-time, asymptotically
non-flat.

The metric is given by \cite{mistaub,kruskal,ntu63}
\begin{eqnarray}
ds^2 &=& -{\cal B}(r) \, \left(dt+2n\cos\theta d\phi\right)^2+ \frac{dr^2}{{\cal B}(r)}+\left(r^2+n^2\right) \left(d\theta^2
+\sin^2\theta d\phi^2 \right) ~.\label{tnn}\\
{\cal B}(r) &=& 1-\frac{2({\cal M}r+n^2)}{r^2+n^2}
\end{eqnarray}
where, ${\cal M}$ denotes the gravito-electric mass or ADM mass  and $n$ denotes
the gravito-magnetic mass or dual mass or magnetic mass of the space-time. The idea
of dual mass  could be found in \cite{sen}. It is evident that there are two type of
singularities are present in the metric (\ref{tnn}). One type occurs at $B(r)=0$ which
gives us the Killing horizons or BH horizons:
\begin{eqnarray}
r_{\pm}= {\cal M}\pm\sqrt{{\cal M}^2+n^2}\,\, \mbox{and}\,\,  r_{+}> r_{-}
\end{eqnarray}
$r_{+}$ is called event horizon (${\cal H}^+$) or outer horizon and $r_{-}$  is called
Cauchy horizon (${\cal H}^{-}$) or inner horizon. The other type of singularity occurs at $\theta=0$
and $\theta=\pi$,  where the determinant of the metric component vanishes. Misner\cite{misner}
showed that in order to remove the apparent singularities at $\theta=0$ and at $\theta=\pi$ , $t$
must be identified modulo $8\pi n$. Provided that $r^2+n^2 \neq 2({\cal M}r+n^2)$.
It should be noted that the NUT parameter actually  measures deviation from the asymptotic
flatness at infinity which could be manifested in the off-diagonal components of the metric and
this is happening due to presence of the Dirac-Misner type of singularity.

As long as
\begin{eqnarray}
{\cal M}^2 + n^2 \geq 0 ~.\label{ineq}
\end{eqnarray}
then the TN metric  describes a BH, otherwise it has a naked  singularity. When ${\cal M}^2 +n^2=0$, 
we find the extremal TN BH.

The product and sum  of horizon radii becomes
\begin{equation}
r_+ r_- = -n^2\,\,\, \mbox{and} \,\, r_{+}+ r_{-} = 2{\cal M} ~.\label{tn3}
\end{equation}

The area\cite{pp15} of this BH is given by
\begin{eqnarray}
{\cal A}_{\pm} &=& \int^{2\pi}_0\int^\pi_0\sqrt{g_{\theta\theta}g_{\phi\phi}}{\mid }_{r=r_{\pm}} d\theta d\phi 
                = 8\pi\left[({\cal M}^2+n^2)\pm {\cal M}\sqrt{{\cal M}^2+n^2}\right]
~.\label{tn}
\end{eqnarray}

Their product\cite{pp15} and sum yields
\begin{eqnarray}
{\cal A}_{+}{\cal A}_{-} &=& (8\pi n)^2\left({\cal M}^2+n^2\right)  \,\,\, \mbox{and} \,\,\, {\cal A}_{+}+{\cal A}_{-} =
16\pi \left({\cal M}^2+n^2\right)   ~. \label{tn1}
\end{eqnarray}
It indicates that the product is dependent on both mass parameter and NUT parameter. Thus it is 
not an universal quantity.  Thus the conjecture ``area product is universal'' does not hold 
for non-asymptotic TN spacetime.

For completeness, we further compute the area minus and area division:
\begin{eqnarray}
{\cal A}_{\pm}- {\cal A}_{\mp} &=& 8\pi {\cal M} T_{\pm} {\cal A}_{\pm} ~.\label{tn2}
\end{eqnarray}
and  
\begin{eqnarray}
\frac{{\cal A}_{+}}{{\cal A}_{-}} &=& \frac{r_{+}^2+n^2}{r_{-}^2+n^2} ~.\label{tn4}
\end{eqnarray}

Again, the sum of area inverse is found to be 
\begin{eqnarray}
\frac{1}{{\cal A}_{+}}+\frac{1}{{\cal A}_{-}} &=& \frac{1}{4\pi n^2}  ~.\label{tn5}
\end{eqnarray}
and the minus of area inverse is computed to be 
\begin{eqnarray}
\frac{1}{{\cal A}_{\pm}}-\frac{1}{{\cal A}_{\mp}} &=& \mp \frac{{\cal M}}{8\pi n^2 \sqrt{{\cal M}^2+n^2}} 
~.\label{tn6}
\end{eqnarray}
It indicates that they are all mass dependent relations.

Likewise, the entropy product\cite{pp15} and entropy sum of ${\cal H}^{\pm}$ becomes:
\begin{eqnarray}
{\cal S}_{-} {\cal S}_{+}  &=& (2\pi n)^2\left({\cal M}^2+n^2\right) \,\,\, \mbox{and} \,\,\,
{\cal S}_{-}+{\cal S}_{+}  = 4\pi\left({\cal M}^2+n^2 \right) ~.\label{tn7}
\end{eqnarray}

For our record, we also compute the entropy minus of ${\cal H}^\pm$ as 
\begin{eqnarray}
{\cal S}_{\pm}- {\cal S}_{\mp} &=&  8\pi {\cal M} T_{\pm} {\cal S}_{\pm} ~.\label{tn8}
\end{eqnarray}
and  the entropy division of ${\cal H}^{\pm}$ as
\begin{eqnarray}
\frac{{\cal S}_{+}}{{\cal S}_{-}} &=& \frac{r_{+}^2+n^2}{r_{-}^2+n^2} ~.\label{tn9}
\end{eqnarray}
Again, the sum of entropy inverse is found to be 
\begin{eqnarray}
\frac{1}{{\cal S}_{+}}+\frac{1}{{\cal S}_{-}} &=& \frac{1}{\pi n^2}  ~.\label{tn10}
\end{eqnarray}
and the minus of entropy inverse is 
\begin{eqnarray}
\frac{1}{{\cal S}_{\pm}}-\frac{1}{{\cal S}_{\mp}} &=& \mp \frac{{\cal M}}{2\pi n^2 \sqrt{{\cal M}^2+n^2}} 
~.\label{tn11}
\end{eqnarray}

The Hawking\cite{bcw73} temperature of ${\cal H}^{\pm}$ reads off
\begin{eqnarray}
T_{\pm} &=& \frac{r_{\pm}-r_{\mp}}{ 8 \pi \left({\cal M}r_{\pm}+n^2 \right)} 
\,\, \mbox{and}\,\, T_{+}> T_{-}.~\label{tn12}
\end{eqnarray}
Their product\cite{pp15} and sum yields
\begin{eqnarray} 
T_{+} T_{-} &=& -\frac{1}{4\pi n^2}
\,\,\, \mbox{and} \,\,\,T_{+} +T_{-} = -\frac{\cal M}{2\pi n^2}  .~\label{tn13}
\end{eqnarray}
It may be noted that surface temperature product is independent of mass while sum depends on 
mass thus the product is universal whereas sum is not universal. It is also shown that for 
TN BH:
\begin{eqnarray}
T_{+}{\cal S}_{+}+T_{-}{\cal S}_{-} &=& 0  .~\label{tn14}
\end{eqnarray}

It is clearly evident that it is a mass independent(universal) relation and implies that 
$T_{+}{\cal S}_{+}= -T_{-}{\cal S}_{-}$ may be taken as a criterion whether there is a 2D CFT 
dual for the BHs in the Einstein gravity and other diffeomorphism gravity theories \cite{castro12,chen12}. 
This universal relation also indicates that the left and right central charges are equal i.e. 
$c_{L}=c_{R}$, which is holographically dual to 2D CFT.

\subsection{Area Bound of TN BH for ${\cal H}^{\pm}$:}
Now we are ready to derive the area bound relations for both the horizons. Using the ineqality Eq. (\ref{ineq}) 
one can obtain ${\cal M}^2 \geq -n^2$. Since  $r_{+} \geq r_{-}$, one can obtain ${\cal A}_{+} \geq {\cal A}_{-} \geq 0$. 
Then the area product gives:
\begin{eqnarray}
{\cal A}_{+}  \geq  \sqrt{{\cal A}_{+} {\cal A}_{-}}=8 \pi n \sqrt{{\cal M}^2+n^2} \geq {\cal A}_{-} 
~.\label{tn15}
\end{eqnarray}
and the area sum gives:
\begin{eqnarray}
 16\pi \left({\cal M}^2+n^2 \right) ={\cal A}_{+}+ {\cal A}_{-} \geq {\cal A}_{+} \geq 
\frac{{\cal A}_{+}+ {\cal A}_{-}}{2}= 8\pi \left({\cal M}^2+n^2 \right) ~.\label{tn16}
\end{eqnarray}
Therefore the area bound for  ${\cal H}^{+}$:
\begin{eqnarray}
8\pi \left({\cal M}^2+n^2 \right)   \leq {\cal A}_{+} \leq 16\pi \left({\cal M}^2+n^2 \right) ~.\label{tn17}
\end{eqnarray}
and  the area bound for  ${\cal H}^{-}$:
\begin{eqnarray}
 0 \leq {\cal A}_{-} \leq 8 \pi n \sqrt{{\cal M}^2+n^2} ~.\label{tn18}
\end{eqnarray}

\subsection{Entropy Bound for ${\cal H}^{\pm}$:}
Similarly, since $r_{+} \geq r_{-}$, one can get ${\cal S}_{+} \geq {\cal S}_{-} \geq 0$. 
Therefore the entropy product gives:
\begin{eqnarray}
{\cal S}_{+}  \geq  \sqrt{{\cal S}_{+} {\cal S}_{-}}=2 \pi n \sqrt{{\cal M}^2+n^2} \geq {\cal S}_{-} 
~.\label{tn19}
\end{eqnarray}
and the entropy sum gives:
\begin{eqnarray}
4\pi \left({\cal M}^2+n^2 \right) ={\cal S}_{+}+ {\cal S}_{-} \geq {\cal S}_{+} \geq 
\frac{{\cal S}_{+}+ {\cal S}_{-}}{2}= 2\pi \left({\cal M}^2+n^2 \right) 
 ~.\label{tn20}
\end{eqnarray}
Thus the entropy bound for  ${\cal H}^{+}$:
\begin{eqnarray}
 2\pi \left({\cal M}^2+n^2 \right) \leq {\cal S}_{+} \leq 4\pi \left({\cal M}^2+n^2 \right)
 ~.\label{tn21}
\end{eqnarray}
and  the entropy bound for  ${\cal H}^{-}$:
\begin{eqnarray}
 0 \leq {\cal S}_{-} \leq 2 \pi n \sqrt{{\cal M}^2+n^2} ~.\label{tn22}
\end{eqnarray}

\subsection{ Irreducible Mass Bound for ${\cal H}^{\pm}$:}

Christodoulou\cite{cd70} had derived a relation between surface area of the ${\cal H}^{+}$ 
and irreducible mass,  which could be written as 
\begin{eqnarray}
{\cal M}_{\text{irr}, \pm} &=& \sqrt{\frac{{\cal A}_{\pm}}{16\pi}}=\sqrt{\frac{{\cal S}_{\pm}}{4\pi}}
~. \label{tn23}
\end{eqnarray}

Now the product and sum  of the irreducible mass for both the horizons are
\begin{eqnarray}
{\cal M}_{\text{irr}, +} {\cal M}_{\text{irr},-} &=&  \frac{n\sqrt{{\cal M}^2+n^2}}{2}  \,\, \mbox{and}\,\,\,
{\cal M}_{\text{irr}, +}^{2} + {\cal M}_{\text{irr},-}^{2}  = {\cal M}^2+n^2
.~\label{tn24}
\end{eqnarray}
From the area bound, we get the irreducible mass bound for TN BH.

For ${\cal H}^{+}$:
\begin{eqnarray}
\frac{\sqrt{{\cal M}^2+n^2}}{\sqrt{2}} \leq {\cal M}_{irr, +} \leq \sqrt{{\cal M}^2+n^2}   ~.\label{tn25}
\end{eqnarray}
and  for ${\cal H}^{-}$:
\begin{eqnarray}
0 \leq {\cal M}_{irr,-} \leq  \frac{\left[n^2({\cal M}^2+n^2)\right]^\frac{1}{4}}
{\sqrt{2}} ~.\label{tn26}
\end{eqnarray}
Eq. \ref{tn25} is nothing but the Penrose inequality, which is the first geometric inequality for 
BHs\cite{peni73}.

\subsection{Temperature Bound for ${\cal H}^{\pm}$:}
In BH thermodynamics, temperature is an important thermodynamic parameter. So there must exists temperature bound
relation on the horizons. As is when  $r_{+} \geq r_{-}$,  one must obtain 
$\mid T_{+} \mid \geq \mid T_{-} \mid \geq 0$.

Therefore the temperature product gives:
\begin{equation}
 | T_{+} | \geq  \sqrt{| T_{+} T_{-}|} = \frac{1}{4\pi n} \geq | T_{-}|  ~.\label{tn27}
\end{equation}

and the temperature sum gives:
\begin{equation}
\frac{{\cal M}}{2\pi n^2} = | T_{+}+ T_{-} | \geq | T_{+} | \geq \frac{| T_{+}+ T_{-} |}{2} 
= \frac{{\cal M}}{4\pi n^2}
 ~.\label{tn28}
\end{equation}
Thus the temperature bound for  ${\cal H}^{+}$:
\begin{equation}
 \frac{{\cal M}}{4\pi n^2} \leq | T_{+} | \leq \frac{{\cal M}}{2\pi n^2}
 ~.\label{tn29}
\end{equation}
and  the temperature bound for  ${\cal H}^{-}$:
\begin{equation}
0 \leq | T_{-} | \leq \frac{1}{4\pi n} ~.\label{tn30}
\end{equation}

\subsection{Bound on Heat Capacity $C_{\pm}$ for ${\cal H}^{\pm}$:}
The specific heat can be defined as 
\begin{eqnarray}
C_{\pm} &=& \frac{\partial{\cal M}}{\partial T_{\pm}} .~\label{c1}
\end{eqnarray}
After some calculations, we obtain the specific heat for both the horizons:
\begin{eqnarray}
C_{\pm} &=& -2\pi (r_{\pm}^2+n^2)  .~\label{c5}
\end{eqnarray}
Their product and sum  on ${\cal H}^{\pm}$ yields:
\begin{eqnarray}
C_{+} C_{-} &=&  (4\pi n)^2 \left({\cal M}^2+n^2 \right)  .~\label{c6}
\end{eqnarray}
and 
\begin{eqnarray}
C_{+}+C_{-} &=&  -8\pi  \left({\cal M}^2+n^2 \right)  .~\label{c7}
\end{eqnarray}
Using ${\cal M}^2 \geq -n^2$ with the product of heat capacity  and the sum of heat capacity, we get 
the bound on heat capacity for both the horizons.
For  ${\cal H}^{+}$:
\begin{eqnarray} 
4\pi  \left({\cal M}^2+n^2 \right) \leq |C_{+}| \leq 8\pi  \left({\cal M}^2+n^2 \right)   ~.\label{c8}
\end{eqnarray}
and for ${\cal H}^{-}$:
\begin{eqnarray}
 0 \leq | C_{-}| \leq 4 \pi n \sqrt{{\cal M}^2+n^2} ~.\label{c9}
\end{eqnarray}
Using Eq. \ref{tn1}, Eq. \ref{tn7} and Eq. \ref{c6}, we find a new relation among 
specific heat product, area product and entropy product which is given by 
\begin{eqnarray}
C_{+} C_{-} &=& \frac{{\cal A}_{+} {\cal A}_{-}}{4} = 4 {\cal S}_{+} {\cal S}_{-}  .~\label{c10}
\end{eqnarray}
So far we have calculated different thermodynamic quantities and these formulae might be 
useful to further understanding the microscopic nature of BH entropy both exterior and interior. 
Again, the entropy product of inner and outer horizons could be used to determine whether the 
classical BH entropy could be written as a Cardy formula, giving some evidence for a holographic 
description of BH/CFT correspondence\cite{chen12,xu15}. The above thermodynamic properties including the 
Hawking temperature and area of both the horizons may therefore be expected to play a crucial role 
in understanding the BH entropy of ${\cal H}^{\pm}$ at the microscopic level.

\section{\label{dis} Conclusion:}
In order to understand the black hole entropy at the microscopic level, we have studied both the 
inner horizon and outer horizon thermodynamics for TN BH in four dimensional Lorentzian geometry. We 
also computed various thermodynamic relations like product, sum, substraction and division of inner and 
outer horizons. Due to the presence of the NUT parameter, we have found that all the thermodynamic relations 
are mass dependent that means they are not universal. 

Using the proposed relations, we also compute different thermodynamic bounds. Like area bound, entropy bound, 
temperature bound etc. We also derived a relation between specific heat product, area product and entropy 
product. The thermodynamic relations that we have derived in this note must have an implications. We suggest 
these thermodynamic formulae gives further clue on understanding the microscopic nature of BH entropy
(both inner and outer) for both the horizons. In \cite{horava15}, we derived thermodynamic product formula for another non-asymptotic type of  BH like Kehagias-Sfetsos BH in Horava-Liftshitz gravity, where the area product is universal in nature.

\appendixname:

In the appendix, we have given various thermodynamic relations for TN BH in comparison with RN BH.

\begin{center}
\begin{tabular}{|c|c|c|}
    \hline
    Parameter($i=+,-$) & RN BH  & Taub-NUT BH\\
    \hline
    $r_{\pm}$: &${\cal M}\pm\sqrt{{\cal M}^{2}-Q^{2}}$ & ${\cal M} \pm\sqrt{{\cal M}^{2}+n^2} $\\

    $\sum r_{i}$: & $2{\cal M}$  & $2{\cal M}$\\

    $\prod r_{i}$: & $Q^2$  & $-n^2$ \\

    ${\cal A}_{\pm}$: & $4\pi \left(2{\cal M}r_{\pm}-Q^2\right)$  & $8\pi \left({\cal M}r_{\pm}+n^2\right)$ \\

    $\sum {\cal A}_{i}$ : & $8\pi\left(2{\cal M}^2-Q^2\right)$ & $16\pi\left({\cal M}^2+n^2\right)$\\

    $\prod {\cal A}_{i}$: &$(4\pi Q^2)^2$  & $(8\pi n)^2\left[{\cal M}^2+n^2\right]$ \\

    ${\cal S}_{\pm}$: &$\pi \left(2{\cal M}r_{\pm}-Q^2\right) $ & $2\pi \left({\cal M}r_{\pm}+n^2 \right)$\\

    $\sum {\cal S}_{i}$ : &$2\pi(2{\cal M}^2-Q^2)$ &  $4\pi\left({\cal M}^2+n^2\right)$\\

    $\prod {\cal S}_{i}$: & $\pi^2Q^4$ & $(2\pi n)^2\left[{\cal M}^2+n^2\right]$\\

  $\kappa_{\pm}$: & $ \frac{r_{\pm}-r_{\mp}}{2 \left(2{\cal M}r_{\pm}-Q^2\right)}$ &  $\frac{r_{\pm}-r_{\mp}}{4 \left({\cal M}r_{\pm}+n^2\right)}$\\

  $\sum{\kappa}_{i}$: &$\frac{4{\cal M}(Q^2-{\cal M}^2)}{Q^4}$ & $-\frac{{\cal M}}{n^2}$\\

  $\prod {\kappa}_{i}$: &$\frac{Q^2-{\cal M}^2}{Q^4}$ & $ -\frac{1}{4n^2}$\\

  $T_{\pm}$: &$\frac{r_{\pm}-r_{\mp}}{4\pi r_{\pm}^2}$ & $ \frac{r_{\pm}-r_{\mp}}{4\pi (r_{\pm}^2+n^2)}$\\

  $\sum T_{i}$:  &$\frac{2{\cal M}(Q^2-{\cal M}^2)}{\pi Q^4} $ & $-\frac{{\cal M}}{2 \pi n^2}$\\

  $\prod T_{i}$ : &$\frac{(Q^2-{\cal M}^2)}{4 \pi^{2}(Q^4)}$ & $-\frac{1}{(4\pi n)^{2}} $\\

  ${\cal M}_{irr, \pm}$: &$\sqrt{\frac{{\cal A}_{\pm}}{16\pi}}$ & $\sqrt{\frac{{\cal A}_{\pm}}{16\pi}}$\\

  $\sum {\cal M}_{irr}^{2}$: & ${\cal M}^2-\frac{Q^{2}}{2}$ & ${\cal M}^2+n^2$\\

  $\prod {\cal M}_{irr}$: &$\frac{Q^2}{4}$& $\sqrt{\frac{n^2({\cal M}^2+n^2)}{4}}$\\
 \hline
\end{tabular}
\end{center}

%\section*{References}

\end{document}